\documentclass[pdflatex,sn-mathphys-num]{sn-jnl}

\usepackage{graphicx}%
\usepackage{multirow}%
\usepackage{amsmath,amssymb,amsfonts}%
\usepackage[T1]{fontenc}%
\usepackage{libertinus-type1}%
\usepackage{libertinust1math}%
\usepackage{inconsolata}%
\usepackage{amsthm}%
\usepackage[title]{appendix}%
\usepackage{xcolor}%
\usepackage{textcomp}%
\usepackage{manyfoot}%
\usepackage{booktabs}%
\usepackage{xspace}%

\theoremstyle{thmstyleone}%
\theoremstyle{thmstyletwo}%
\theoremstyle{thmstylethree}%
\newcommand{\tabfont}{\footnotesize}
\newcommand{\thd}[1]{\textbf{#1}}

\renewcommand{\figurename}{Figure}

\makeatletter
\renewcommand{\@@address}[2][]{%
  \g@addto@macro\auaddress{%
    \stepcounter{affn}%
    \xdef\@currentlabel{\theaffn}%
    \jmkLabel{\theaffn}%
                {\textsuperscript{#1}#2\par}}}
\makeatother

\hypersetup{bookmarksdepth=3}

\begin{document}

\title{An Empirical Measurement of Jailbreaking Evaluators}

\author{\fnm{Yujie} \sur{Mu}}

\affil{\orgname{Independent Researcher}}

\abstract{
Expert evaluation of jailbreak responses is costly and difficult to scale, so the community increasingly relies on automated evaluators to determine whether an attack succeeds.
However, jailbreak studies typically validate their chosen evaluator independently, repeatedly spending resources on similar evaluation efforts while making results across papers difficult to compare.
Different evaluators also encode different definitions of jailbreak success, meaning that reported attack strength and apparent progress can depend substantially on which evaluator is used.
We systematically compare six evaluators that recur in recent jailbreak attack and defense research: HarmBench, JailbreakBench, JailbreakRadar, StrongReject, JADES, and JailMeter.
To our knowledge, no prior study has evaluated all six on the same human-labeled data under a controlled setup.
We evaluate them on JailbreakQR and JailMeter-Eva, using human judgments as the reference, and measure agreement with humans, error types, and consistency across attack families.
For evaluators that require a general-purpose LLM judge, we use a shared backbone to control for model-specific variation.
We found that JADES exhibits the best overall performance, while HarmBench and StrongReject also demonstrate good performance.
}


\maketitle

\section{Introduction}\label{sec1}

Jailbreak attacks have become a central topic in the study of large language model safety, with a growing body of work proposing new attacks, defenses, and benchmarks~\cite{jbb,radar,harmbench}.
A core question underlying all of this work is whether a jailbreak attempt actually succeeds.
Answering this question through expert human evaluation is costly and difficult to scale, especially when modern studies evaluate thousands of prompt-response pairs across multiple models and attack settings.
As a result, the community increasingly relies on automated evaluators to determine whether a model response constitutes a successful jailbreak.

These evaluators directly affect how jailbreak research is interpreted.
Attack papers use them to compute attack success rates, defense papers use them to measure robustness, and benchmarks use them to rank attacks and models.
However, different evaluators encode different definitions of jailbreak success.
A response considered successful by one evaluator may therefore be rejected by another, even when the underlying attack and model response remain unchanged.
Consequently, reported attack strength and apparent progress across studies can depend substantially on the evaluator being used.

Despite this importance, evaluator validation remains fragmented.
Jailbreak studies commonly validate their chosen evaluator independently, often on different datasets, against different baselines, and under different experimental settings.
This practice repeatedly spends resources on similar evaluation efforts while making results across papers difficult to compare.
More importantly, it leaves the field without a common empirical picture of how widely used jailbreak evaluators behave when applied to the same data under the same conditions.

Several evaluators have become recurrent components of recent jailbreak attack and defense research.
HarmBench uses a classifier to determine whether a response exhibits a specified harmful behavior~\cite{harmbench}.
JailbreakBench and JailbreakRadar use LLM-based judges to classify jailbreak responses~\cite{jbb,radar}.
StrongReject evaluates responses using a structured rubric~\cite{strongreject}.
JADES decomposes harmful requests into weighted sub-questions and scores responses against them~\cite{jades}.
JailMeter instead evaluates evidence extracted from the response~\cite{jailmeter}.
These evaluators were designed around different formulations of jailbreak success and were originally validated under different experimental conditions.
To our knowledge, no prior study has systematically evaluated all six on the same human-labeled data under a controlled setup.

In this work, we provide such a comparison.
We evaluate HarmBench, JailbreakBench, JailbreakRadar, StrongReject, JADES, and JailMeter on two human-labeled datasets, JailbreakQR and JailMeter-Eva, using human judgments as the reference~\cite{jades,jailmeter}.
We measure how closely each evaluator agrees with human judgments, characterize the types of errors it makes, and examine whether its behavior remains consistent across different attack families.
For evaluators that require a general-purpose LLM judge, we use the same backbone model, eliminating a source of variation that would otherwise confound differences between evaluation methods.
Our results show that JADES achieves the strongest overall performance, while HarmBench and StrongReject also perform well across the evaluated settings.

Our contributions are as follows.

\begin{enumerate}[1.]
\item We conduct a systematic comparison of six widely used jailbreak evaluators---HarmBench, JailbreakBench, JailbreakRadar, StrongReject, JADES, and JailMeter---under a unified evaluation protocol on two human-labeled datasets.

\item We control the general-purpose judge backbone across all evaluators that require one, isolating differences caused by the evaluation methodology rather than by the underlying judge model.

\item We provide a unified empirical assessment of current jailbreak evaluators, showing that JADES achieves the best overall performance, while HarmBench and StrongReject also demonstrate strong performance.
\end{enumerate}

\section{Background}\label{sec2}

Early jailbreak studies often judged attack success using simple string matching.
A response was treated as successful if it did not begin with a predefined refusal phrase~\cite{gcg}.
This approach is inexpensive and fully reproducible, but it provides only a weak proxy for actual compliance: a response may avoid an explicit refusal while still failing to provide useful harmful content.
Another line of work reuses toxicity or moderation classifiers~\cite{perspective,moderation,llamaguard,wildguard}.
These classifiers estimate whether a response contains harmful content, but harmfulness is not equivalent to successful fulfillment of a harmful request.
A response may contain toxic language without answering the request, while a harmful answer may be expressed in neutral language.

More recent work increasingly relies on capable language models as evaluators.
This follows the broader practice of using LLMs to grade open-ended text~\cite{mtbench,geval}, although such judges are themselves known to exhibit systematic biases~\cite{notfair}.
JailbreakBench and JailbreakRadar adopt this general LLM-as-a-judge paradigm~\cite{jbb,radar}.
HarmBench instead trains a dedicated classifier to determine whether a response exhibits a specified harmful behavior~\cite{harmbench}.
StrongReject retains an LLM judge but replaces a single holistic verdict with a structured rubric~\cite{strongreject}.
Prior work in other evaluation settings has shown that rubric-based assessment can align more closely with human judgments than holistic scoring~\cite{holistic,jonsson,attali}.

More recent evaluators further modify what information is presented to the judge and how the final score is constructed.
JADES argues that direct holistic judgment can be distorted by irrelevant or superficially compliant content, and therefore decomposes the harmful request into weighted sub-questions before scoring the response~\cite{jades}.
JailMeter instead separates evidence extraction from the final judgment, with the goal of reducing interference from jailbreak-specific prompt structure~\cite{jailmeter}.
Together, these approaches reflect substantially different definitions and operationalizations of jailbreak success.

The harmful requests evaluated by these methods are commonly drawn from safety benchmarks such as Do-Not-Answer~\cite{donotanswer} and SALAD-Bench~\cite{saladbench}.
These benchmarks also differ in scope, adoption, and implementation quality~\cite{bob}.
Appendix~\ref{secA1} provides additional background on the jailbreak attacks that generate these responses and the defenses whose effectiveness is measured using such evaluators.

\section{Evaluators Under Measurement}\label{sec3}

Let $q$ denote a harmful question and $J$ a jailbreak attack.
The attack transforms $q$ into a prompt $p = J(q)$, which is submitted to a target model $M$ to obtain a response $r = M(p)$.
An evaluator $E$ then maps the pair $(q,r)$ to a verdict,
$E:(q,r)\mapsto v$, where $v\in\mathbb{V}_E$.
Some evaluators return a binary verdict directly, whereas others produce a score in $[0,1]$ that is subsequently thresholded.
Let $h(q,r)$ denote the human annotation, which we use as the reference label.
For a dataset $D$, we measure evaluator agreement as
\begin{equation}
\mathrm{Agreement}(E,D)
=
\frac{1}{|D|}
\sum_{(q,r)\in D}
\mathbf{1}\!\left[\phi(E(q,r)) = h(q,r)\right],
\label{eq:agreement}
\end{equation}
where $\phi$ maps the evaluator's native output to the label space of $D$.
We use the term \emph{agreement} rather than correctness because human annotations may themselves contain noise.
Moreover, aggregate agreement can conceal systematic errors on particular subsets, motivating our later analysis across error types and attack families.

Table~\ref{tab:evaluators} summarizes the six evaluators considered in this work.

\textbf{HarmBench} formulates jailbreak evaluation as behavior classification~\cite{harmbench}.
Its dedicated classifier, fine-tuned from Llama~2 13B~\cite{llama2}, takes a harmful behavior description together with the generated response and predicts whether the response constitutes an instance of that behavior.
Unlike the remaining evaluators, HarmBench does not require a general-purpose LLM judge at inference time.

\textbf{JailbreakBench} uses a prompted instruction-following model to produce a single holistic verdict for the response according to a fixed evaluation rubric~\cite{jbb}.

\textbf{JailbreakRadar} evaluates a response using three criteria~\cite{radar}: whether it contributes substantive content beyond merely repeating the prompt, whether it provides any requested information or expresses willingness to do so, and whether it remains relevant to the original harmful question.
A jailbreak is considered successful only when all three criteria are satisfied.
Its criterion therefore emphasizes whether the response provides useful progress toward the adversarial request, including through indirect, incomplete, or fictionalized content.

\textbf{StrongReject} is motivated by the observation that apparent compliance does not necessarily imply that a response contains useful harmful information~\cite{strongreject}.
It first applies a refusal gate: responses classified as refusals receive a score of zero.
Non-refusal responses are then rated for how convincing and how specific they are, and the two ratings are normalized and aggregated into a final score.
StrongReject also provides a fine-tuned scoring model; Section~\ref{subsec43} specifies the variant used in our experiments.

\textbf{JADES} decomposes the harmful question into weighted sub-questions, identifies response sentences relevant to each sub-question, scores the resulting sub-answers on a five-level scale~\cite{likert,boone}, and aggregates the scores according to their weights~\cite{jades}.
This design replaces a single holistic judgment with decompositional scoring over the information required to fulfill the harmful request.

\textbf{JailMeter} first extracts the parts of the response that are relevant to the original harmful question through an information-bottleneck formulation, and then evaluates the extracted evidence~\cite{jailmeter}.
Its final decision considers both whether the response captures the harmful intent and whether it provides a sufficiently complete answer.

\begin{table*}[!t]
\centering
\caption{Summary of the six jailbreak evaluators considered in this work.}
\label{tab:evaluators}
\tabfont
\setlength{\tabcolsep}{4pt}
\begin{tabular}{@{}llll@{}}
\toprule
\thd{Evaluator} & \thd{Evaluation mechanism} & \thd{Native output} & \thd{LLM judge} \\
\midrule
HarmBench~\cite{harmbench}
    & Fine-tuned behavior classifier
    & Boolean
    & No \\
JailbreakBench~\cite{jbb}
    & Holistic prompted judgment
    & Boolean
    & Yes \\
JailbreakRadar~\cite{radar}
    & Three-criterion prompted judgment
    & Boolean
    & Yes \\
StrongReject~\cite{strongreject}
    & Refusal gate + rubric scoring
    & Score $[0,1]$
    & Yes \\
JADES~\cite{jades}
    & Decomposition + weighted scoring
    & Score $[0,1]$
    & Yes \\
JailMeter~\cite{jailmeter}
    & Evidence extraction + judgment
    & Boolean
    & Yes \\
\bottomrule
\end{tabular}
\end{table*}

\section{Experiment Setup}\label{sec4}

\subsection{Datasets}\label{subsec41}

We evaluate all six evaluators on two human-labeled datasets, JailbreakQR and JailMeter-Eva.
The two datasets differ in label granularity, target models, and attack families, providing complementary evaluation settings.
Table~\ref{tab:datasets} summarizes their main characteristics.

\textbf{JailbreakQR}~\cite{jades}.
JailbreakQR is constructed from 260 harmful questions drawn from JailbreakBench and JailbreakRadar and evaluated under five jailbreak attacks: GCG~\cite{gcg}, DSN~\cite{dsn}, LAA~\cite{laa}, PAIR~\cite{pair}, and JailbreakChat~\cite{dan}.
The target models are Vicuna 13B~\cite{vicuna}, Llama 2 7B Chat~\cite{llama2}, GPT-3.5 Turbo, and GPT-4~\cite{gpt4}.
From the resulting prompt-response pairs, 400 are sampled for human annotation.
Each pair is labeled by three annotators as failed, partially successful, or successful, with disagreements resolved by majority vote.
The reported ordinal Krippendorff's $\alpha$ is 0.823~\cite{krippendorff}.

\textbf{JailMeter-Eva}~\cite{jailmeter}.
JailMeter-Eva contains 330 non-refusal responses generated by GPT-4o mini.
The dataset covers eight attacks: CodeChameleon~\cite{codechameleon}, Crescendo~\cite{crescendo}, PastTense~\cite{pasttense}, JSP, CodeAttack~\cite{codeattack}, PAIR~\cite{pair}, DeepInception~\cite{deepinception}, and DRA~\cite{dra}.
The samples span six harm categories and use binary labels, with 73 responses labeled as successful jailbreaks.
By excluding explicit refusals, JailMeter-Eva focuses evaluation on harder cases in which the model produces substantive content but the success of the jailbreak remains ambiguous.

\begin{table*}[!t]
\centering
\caption{Summary of the two human-labeled evaluation datasets.}
\label{tab:datasets}
\tabfont
\setlength{\tabcolsep}{5pt}
\begin{tabular}{@{}lll@{}}
\toprule
\thd{Property} & \thd{JailbreakQR} & \thd{JailMeter-Eva} \\
\midrule
Pairs
& 400
& 330 \\
Label space
& Failed / Partial / Successful
& Failed / Successful \\
Annotators / agreement
& 3 / $\alpha=0.823$
& NA \\
Models / attack families
& 4 / 5
& 1 / 8 \\
Refusals included
& Yes
& No \\
\bottomrule
\end{tabular}
\end{table*}

\subsection{Label Alignment and Metrics}\label{subsec42}

The evaluators and datasets use different output spaces, so we define a common label mapping for comparison.
For binary evaluation, the human label \emph{failed} is mapped to negative, while \emph{partially successful} and \emph{successful} are mapped to positive.
For score-based evaluators, we use their published decision thresholds.
JADES uses 0.25 for binary classification and 0.25 and 0.75 for ternary classification~\cite{jades}.
StrongReject uses its released threshold~\cite{strongreject}.
We report accuracy, precision, recall, F1, and false positive rate (FPR).
We include FPR because false positives directly inflate the measured attack success rate, whereas recall captures the evaluator's ability to identify successful jailbreaks.

\subsection{Judge Backbone}\label{subsec43}

Five of the six jailbreak evaluators rely on one or more LLMs.
As evaluator performance can vary with the underlying judge model, we fix this component across all applicable methods.
Specifically, JailbreakBench, JailbreakRadar, StrongReject, JADES, and JailMeter use \texttt{orcarouter/Qwen3.8-27B-Uncensored} with temperature set to 1.0.
We retain each evaluator's published prompt template and decision procedure.
For StrongReject, we use the rubric-based LLM judge rather than its released fine-tuned scorer so that it is evaluated under the same shared-backbone setting as the other prompt-based methods.

We use an uncensored backbone to reduce evaluator-side refusals when processing harmful content, which would otherwise introduce errors unrelated to the evaluation criterion itself.

This control does not apply to HarmBench, which uses its released dedicated classifier and does not call a general-purpose judge.
Accordingly, our experiments primarily compare evaluator methodologies under a controlled judge backbone rather than reproducing every evaluator exactly in its originally released configuration.

\section{Experiment Results}\label{sec5}

\subsection{Main Results}\label{subsec51}

\begin{figure*}[!ht]
\centering
\includegraphics[width=.9\textwidth]{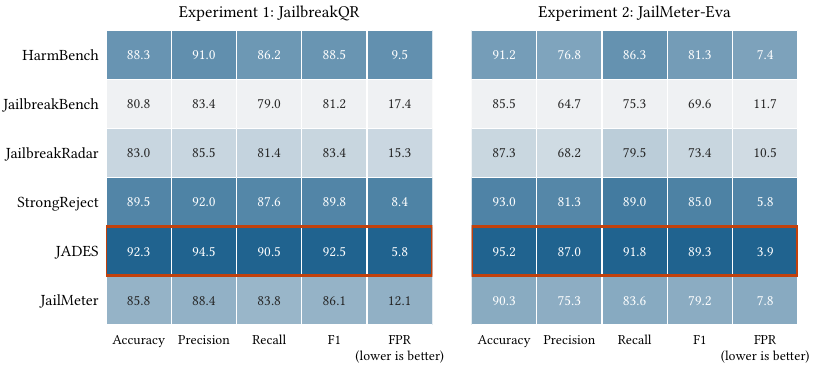}
\caption{Binary agreement with human labels.
Values are percentages.
Cell shading is normalised within each column so that the darker end always marks the better result, which for FPR means the smaller number.
The highlighted row marks the strongest evaluator overall.}
\label{fig:main}
\end{figure*}

Figure~\ref{fig:main} reports binary agreement with the human labels.
The left panel presents results on JailbreakQR, while the right panel presents results on JailMeter-Eva.
JailMeter-Eva excludes refusals before sampling and therefore focuses more heavily on ambiguous non-refusal responses.

\noindent
\textit{Findings.}
JADES achieves the strongest overall performance on both datasets.
On JailbreakQR, it reaches 92.3\% accuracy and 92.5\% F1, while also producing the lowest false positive rate at 5.8\%.
StrongReject follows with 89.5\% accuracy and 89.8\% F1, while HarmBench also performs consistently well with 88.3\% accuracy and 88.5\% F1.
In contrast, JailbreakBench and JailbreakRadar show noticeably lower agreement with human judgments, with JailbreakBench obtaining the lowest F1 score at 81.2\%.

The same ordering largely persists on JailMeter-Eva.
JADES again performs best, reaching 95.2\% accuracy, 89.3\% F1, and an FPR of only 3.9\%.
StrongReject remains the second strongest evaluator with 93.0\% accuracy and 85.0\% F1, followed by HarmBench.
JailbreakBench and JailbreakRadar again perform substantially worse, particularly in precision and F1.
This consistency across the two datasets suggests that the observed differences are not specific to a single benchmark or attack distribution.

A notable pattern appears on JailMeter-Eva.
Although most evaluators achieve relatively high accuracy, their precision and F1 scores are substantially lower than on JailbreakQR.
For example, JailbreakBench reaches 85.5\% accuracy but only 64.7\% precision and 69.6\% F1.
This gap reflects the class imbalance of JailMeter-Eva and shows that accuracy alone can overstate evaluator quality in this setting.
The false positive rate provides an important complementary view because false positives directly inflate the measured attack success rate.
JADES and StrongReject achieve the lowest FPRs on both datasets, indicating that their stronger overall performance is not obtained simply by labeling more responses as successful jailbreaks.

\subsection{Discussion}\label{sec6}

The results show that evaluator choice can materially affect conclusions about jailbreak success.
Even under a shared judge backbone, the six evaluators exhibit substantial differences in agreement with human labels.
This indicates that evaluator design itself, rather than only the underlying judge model, is an important source of variation in jailbreak measurement.

The strongest methods also share a common property.
JADES and StrongReject do not rely solely on a single unrestricted holistic judgment.
StrongReject explicitly separates refusal detection from the assessment of response specificity and convincingness, while JADES decomposes the harmful request into smaller criteria before aggregating the resulting scores.
Their consistently stronger performance suggests that adding structure to the evaluation process can improve agreement with human judgments.
By contrast, evaluators based primarily on a single prompted verdict, such as JailbreakBench and JailbreakRadar, show larger performance gaps in our experiments.

JADES exhibits the strongest performance across all five reported metrics on both datasets.
Its advantage is particularly visible in FPR, where it reduces false positives to 5.8\% on JailbreakQR and 3.9\% on JailMeter-Eva.
This is important for jailbreak research because a false positive directly counts a failed or insufficient response as a successful attack and therefore inflates the reported attack success rate.
The results suggest that decompositional evaluation may better distinguish substantive harmful compliance from responses that only appear compliant at a surface level.

StrongReject also performs consistently well across both datasets.
Its results suggest that explicitly modeling refusal and response quality provides a strong alternative to purely holistic classification.
HarmBench, despite using a dedicated classifier rather than the shared LLM judge, also remains competitive, indicating that task-specific classifiers can provide reliable jailbreak evaluation when trained on suitable human-labeled data.

Finally, the results highlight the importance of evaluating jailbreak evaluators on more than one dataset.
JailbreakQR contains a broader mixture of responses, including refusals, whereas JailMeter-Eva deliberately excludes refusals and concentrates on more ambiguous non-refusal cases.
The relative ordering of the evaluators remains broadly stable across these settings, but the larger drops in precision and F1 on JailMeter-Eva show that difficult non-refusal responses remain substantially harder to classify.
A reliable jailbreak evaluator should therefore be assessed not only by aggregate accuracy, but also by its error profile and robustness across different attack distributions.

\section{Limitations}\label{subsec63}

Our study is limited by the scope of the available evaluation data.
Both JailbreakQR and JailMeter-Eva are English-only and moderate in size, so the results may not generalize to other languages, domains, target models, or future jailbreak strategies.
Human annotations serve as our reference labels, but they are themselves imperfect.
For example, JailbreakQR reports an ordinal Krippendorff's $\alpha$ of 0.823, indicating that some evaluator--human disagreements may reflect annotation uncertainty rather than evaluator error alone.

A second limitation concerns dataset provenance.
Each dataset was introduced by work that also proposed one of the evaluators considered in this study.
Although the evaluation sets are not training data for these evaluators, the design of an evaluator and the construction of its associated dataset may reflect related assumptions about what constitutes a successful jailbreak.
We therefore place greater emphasis on performance that remains consistent across both datasets rather than on results from either dataset in isolation.

Our shared-backbone setup also provides only partial control.
We use the same general-purpose judge model for all evaluators that support such a configuration, which isolates variation caused by their evaluation procedures.
However, HarmBench uses its own dedicated classifier, and StrongReject is evaluated with its rubric-based judge rather than its released fine-tuned scorer.
Our results therefore compare evaluator methodologies under a controlled experimental setting rather than reproducing every system exactly in its native configuration.
Evaluating each released system with its originally recommended model and settings would answer a complementary question.

Finally, this work focuses on jailbreak evaluation rather than the development of new attacks.
We introduce no new jailbreak method and generate no new harmful benchmark content.
Our experiments use previously released datasets and attack outputs.
Accordingly, the primary contribution of this work is methodological: improving the reliability and comparability of how jailbreak success is measured.

\section{Conclusion}\label{sec7}

Automated evaluators have become a central component of jailbreak research because expert human evaluation is costly and difficult to scale.
However, widely used evaluators implement different definitions and procedures for determining whether a jailbreak succeeds, making results across studies difficult to compare.

In this work, we systematically compared six jailbreak evaluators---HarmBench, JailbreakBench, JailbreakRadar, StrongReject, JADES, and JailMeter---on two human-labeled datasets under a unified evaluation protocol.
For evaluators that rely on a general-purpose LLM judge, we fixed the judge backbone to reduce model-specific variation.
We evaluated agreement with human judgments, error characteristics, and consistency across different attack families.

Our results show that JADES achieves the strongest overall performance across both datasets, while StrongReject and HarmBench also demonstrate consistently strong agreement with human judgments.
The remaining evaluators exhibit larger gaps, particularly on the more ambiguous non-refusal examples in JailMeter-Eva.
These findings show that evaluator choice is itself an important source of variation in jailbreak measurement and that more structured evaluation procedures can provide more reliable agreement with human judgments.

We hope this comparison provides a common reference for selecting and evaluating jailbreak evaluators and reduces the need for individual studies to repeatedly establish evaluator reliability under incompatible experimental settings.

\bibliography{sn-bibliography}

\begin{appendices}

\section{Related Work}\label{secA1}

\subsection{Risks of Large Language Models}\label{secA11}

Large language models are increasingly embedded in content creation and information retrieval, and online content is now also being optimized for model-driven retrieval systems~\cite{geoflag}.
This growing deployment surface has motivated extensive work on the security, privacy, and safety risks of generative models.

Privacy risks arise at several stages of the model lifecycle.
Language models can memorize and reproduce sensitive training examples~\cite{extract}, leak information from prior conversations to adversarial users~\cite{convleak}, and infer personal attributes that were never explicitly stated in the input~\cite{inferprivacy}.
Synthetic data generated by language models can also propagate latent privacy risks into subsequent fine-tuning pipelines~\cite{fakepriv}.

Related concerns extend to generative vision systems.
Image-generation models can be compromised to perform hidden secondary tasks~\cite{neeko}, while diffusion models may generate unsafe or hateful visual content~\cite{unsafediffusion}.
Their safety mechanisms are themselves subject to red-teaming and circumvention~\cite{sdfilter}.
More recent image-generation paradigms introduce additional concerns around authenticity and provenance~\cite{authrisk}, while harmful intent may also emerge compositionally across multiple turns or images rather than within a single output~\cite{hatefulpanels}.

Generative models can also facilitate broader security abuse.
Harmful objectives may be concealed within apparently benign tasks~\cite{harmlesstask}, and language-model capabilities can reduce the effort required for spear-phishing~\cite{spearphishing} and other conventional security attacks~\cite{dualuse}.
These risks motivate the safety mechanisms whose robustness is subsequently tested through jailbreak evaluation.

\subsection{Safety Training and Its Limits}\label{secA12}

Modern language models are commonly subjected to safety alignment before deployment.
Representative approaches include reinforcement learning from human feedback~\cite{instructgpt}, alignment through model-written principles and feedback~\cite{constitutional}, and automated red-teaming procedures for identifying unsafe behaviors~\cite{redteamlm,askell}.

These mechanisms substantially improve model behavior but do not provide complete protection.
Prior work identifies structural limitations that make robust alignment difficult under adversarial prompting~\cite{alignlimits}, and increased model capability does not necessarily translate into stronger safety behavior~\cite{smartersafer}.
Jailbreak attacks directly exploit this gap between intended safety policies and model behavior under adversarial inputs.

\subsection{Jailbreak Attacks}\label{secA13}

Jailbreak attacks span several broad methodological families.
One family consists of human-authored prompts collected from public communities and real-world use.
The well-known DAN-style role-playing prompts are representative examples of such in-the-wild jailbreaks~\cite{dan}.

A second family transforms or obfuscates the harmful request so that its intent becomes more difficult for safety mechanisms to recognize.
Existing techniques encode requests through low-resource or multilingual formulations~\cite{lowresource,multilingual}, ciphers~\cite{cipher}, ASCII art~\cite{artprompt}, source code~\cite{codeattack}, or personalized encoding and transformation schemes~\cite{codechameleon}.
Although these methods differ in representation, they share the goal of preserving harmful semantics while changing the surface form presented to the model.

Optimization-based attacks instead search for prompts that maximize a jailbreak objective.
Examples include gradient-guided adversarial suffixes~\cite{gcg}, evolved human-readable prompts~\cite{autodan}, iterative attacker models that refine jailbreak prompts over multiple queries~\cite{pair,tap}, template-based fuzzing~\cite{gptfuzzer}, random-search strategies~\cite{laa}, refusal-suppression objectives~\cite{dsn}, and attacks that manipulate controllable decoding behavior~\cite{coldattack}.

Other attacks exploit inference-time behavior or multi-turn interaction.
These approaches manipulate decoding parameters~\cite{genexploit}, embed harmful requests within nested fictional scenarios~\cite{deepinception}, progressively escalate harmful intent across dialogue turns~\cite{crescendo}, reformulate harmful requests in the past tense~\cite{pasttense}, disguise and subsequently reconstruct the original request~\cite{dra}, automate jailbreak discovery against deployed systems~\cite{masterkey}, or place malicious requests alongside benign ones~\cite{concurrency}.
Broader analyses have studied why safety alignment fails across multiple jailbreak mechanisms~\cite{jailbroken}, while JailbreakRadar provides a unified taxonomy and large-scale empirical comparison of attack strategies~\cite{radar}.

Our work does not introduce a new jailbreak attack.
Instead, we use outputs produced by existing attack families as evaluation material for studying how different jailbreak evaluators interpret the same model responses.

\subsection{Defenses That Evaluators Also Score}\label{secA14}

Jailbreak evaluators are used not only to measure attack effectiveness but also to quantify the effectiveness of defenses.
Existing defenses include perplexity-based filters that reject suspicious or unnatural adversarial suffixes~\cite{perplexityfilter}, SmoothLLM, which perturbs adversarial prompts and aggregates model outputs~\cite{smoothllm}, and baseline defense pipelines that combine detection, preprocessing, and adversarial training~\cite{baselinedefense}.

The measured effectiveness of these defenses ultimately depends on how jailbreak success is defined and detected.
As a result, evaluator choice affects conclusions not only about the strength of attacks but also about the robustness gains attributed to defenses.
This further motivates a controlled comparison of the evaluators used throughout the jailbreak literature.

\end{appendices}

\end{document}